\documentclass{article}
\usepackage[T1]{fontenc} % add special characters (e.g., umlaute)
\usepackage[utf8]{inputenc} % set utf-8 as default input encoding
\usepackage{ismir,amsmath,cite,url}
\usepackage{amsmath}
\usepackage{graphicx}
\usepackage{color}
\usepackage{amssymb}% http://ctan.org/pkg/amssymb
\usepackage{pifont}% http://ctan.org/pkg/pifont
\newcommand{\cmark}{\ding{51}}%
\newcommand{\xmark}{\ding{55}}%
\usepackage{adjustbox}
\usepackage{booktabs}
\usepackage[table]{xcolor}
\usepackage{comment}
\usepackage{stfloats}
\usepackage{siunitx}
\usepackage[bookmarks=false]{hyperref}

\usepackage{lineno}
\title{GOAT: A Large Dataset of Paired Guitar Audio Recordings and Tablatures}

 \multauthor
 {Jackson Loth\textsuperscript{1}, Pedro Sarmento\textsuperscript{1,2}, Saurjya Sarkar\textsuperscript{1}, Zixun Guo\textsuperscript{1}, Mathieu Barthet\textsuperscript{1,3},} {and Mark Sandler\textsuperscript{1}} {\\\textsuperscript{1}Centre for Digital Music,
  Queen Mary University of London\\
  \textsuperscript{2}Music.AI\\
  \textsuperscript{3}Aix-Marseille Univ CNRS PRISM\\
 {\tt\small \{j.j.loth, p.p.sarmento, saurjya.sarkar, zixun.guo, m.barthet,	mark.sandler\}@qmul.ac.uk}
 }

 \def\authorname{J. Loth, P. Sarmento, S. Sarkar, Z. Guo, M. Barthet, and M. Sandler}
\begin{document}

\maketitle

\begin{abstract} 
In recent years, the guitar has received increased attention from the music information retrieval (MIR) community driven by the challenges posed by its diverse playing techniques and sonic characteristics. Mainly fueled by deep learning approaches, progress has been limited by the scarcity and limited annotations of datasets. To address this, we present the Guitar On Audio and Tablatures (GOAT) dataset,  comprising 5.9 hours of unique high-quality direct input audio recordings of electric guitars from a variety of different guitars and players. We also present an effective data augmentation strategy using guitar amplifiers which delivers near-unlimited tonal variety, of which we provide a starting 29.5 hours of audio. Each recording is annotated using guitar tablatures, a guitar-specific symbolic format supporting string and fret numbers, as well as numerous playing techniques. For this we utilise both the Guitar Pro format, a software for tablature playback and editing, and a text-like token encoding. Furthermore, we present competitive results using GOAT for MIDI transcription and preliminary results for a novel approach to automatic guitar tablature transcription. We hope that GOAT opens up the possibilities to train novel models on a wide variety of guitar-related MIR tasks, from synthesis to transcription to playing technique detection.
\end{abstract}
\section{Introduction}

The guitar is one of the most popular instruments in modern western music. It is no surprise that guitar-centered research has received a lot of attention, particularly in the wake of advances in deep learning. Tasks such as automatic guitar transcription (AGT) \cite{wiggins2019guitar} \cite{riley2024high} \cite{cwitkowitz2023fretnet} \cite{riley2024gaps}, tablature generation \cite{loth2023proggp} \cite{sarmento2023shredgp}, 
guitar amplifier modeling \cite{wright_real-time_2020} \cite{wright_adversarial_2022} \cite{chen2024towards}, and more have seen great progress. Despite this interest in guitars, there remains limited data availability for guitar audio, particularly in the case of annotated audio recordings. 

Musical Instrument Digital Interface (MIDI) is a well known standard of representing music, and is a common choice when annotating the guitar \cite{chen2022towards} \cite{riley2024gaps}. MIDI easily encodes essential parameters of musical notes such as pitch, note onset, duration and velocity. However, it does not give any indication of string or fret number, nor does it have any standardized way of representing the numerous expressive components of guitar playing \cite{loth2024midi}. It is based on a descriptive type of notation, in which there is a relationship between symbols and pitch. Tablatures, however, are a prescriptive type of notation commonly used for guitar, in which there is a relation between symbols and actions, i.e. how to play said symbols \cite{sarmento_dadagp_2021}. While there are some guitar-focused datasets in the literature that offer tablature-like annotations \cite{xi_guitarset_2018} \cite{zang2024synthtab} \cite{abreu2024leveraging} \cite{pedroza2025guitar}, they are typically limited to fret and string values, ignoring expressive techniques (e.g. bends, palm mutes, legatos).

As a motivation behind the creation of GOAT, the \textbf{G}uitar \textbf{O}n \textbf{A}udio and \textbf{T}ablatures dataset, we seek to surpass the limitations of MIDI and instead prioritize building a dataset of guitar recordings annotated using guitar tablatures, a popular musical format among guitarists. For this we rely on Guitar Pro, a commercial software for tablature edition and playback widely spread amongst the guitar community. This dataset consists of 5.9 hours of unique real audio recordings of guitars, fully annotated for fret/string numbers and expressive playing techniques. We also include the entire dataset rendered through a wide variety of different digital guitar amplifiers in various configurations and cabinet impulse responses, for a total of 29.5 hours of fully annotated audio. The annotations are provided in Guitar Pro format, as well as DadaGP \cite{sarmento_dadagp_2021} tokens, a compressed text-like representation of the information in the tablatures. In an attempt to make a bridge with prior literature, we also include MIDI versions of the content. 

This paper's contributions include (1) an overview of the GOAT dataset and the data collection methodology;
(2) a data augmentation strategy for obtaining a large amount of timbral variation; (3) an evaluation of results when using the GOAT dataset for the task of AGT, complemented with (4) preliminary results on a novel audio-to-text-based approach for AGT; finally, some (5) suggestions for prospective applications using the GOAT dataset. We first cover some relevant background concerning guitar tablatures and similiar previously released music datasets. In Section \ref{sec:met}, we extensively describe the methodology used to compile the GOAT dataset. We then describe, in Section \ref{sec:dataset}, the details of the GOAT dataset, the features it encompasses and the ones it lacks. Within Section \ref{sec:experiments} we present results from a use case for GOAT in the context of AGT, following both a traditional MIDI-based approach for comparison with previous results, and a novel procedure using the text-like token representations of GOAT to fine-tune a Whisper \cite{radford2023robust} model. Finally, Section \ref{sec:app} proposes additional applications for the dataset. 
\section{Background}

\subsection{Guitar Tablatures}

\begin{figure}[h]
    \centering
    \includegraphics[width=\columnwidth]{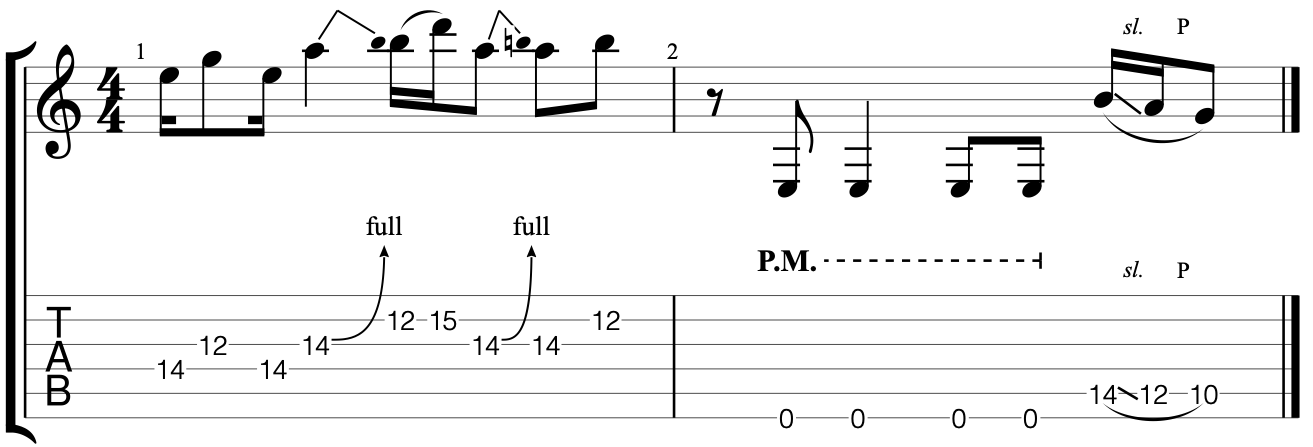}
    \caption{Example of two measures of a guitar tablature, from the Guitar Pro editing software, exemplifying different guitar playing techniques (i.e. bends, palm mutes, legatos and slides).}
    \label{fig:tabexample}
\end{figure}

Guitar tablatures (see Figure \ref{fig:tabexample}), also called tabs, are symbolic representations of guitar music.

In contrast to MIDI, which simply represents a note pitch over time, tabs represent both the fret and string number of a guitar. Tablatures can also support expressive playing techniques such as bends, hammer-ons, pull-offs, strum directions and more. It is a very intuitive, visual representation of guitar music and is popular among guitarists for learning specific songs and performances. Guitar tabs have seen an increase in attention from researchers in the past few years in areas such as guitar tablature generation \cite{sarmento_dadagp_2021} \cite{loth2023proggp} \cite{sarmento2023shredgp} \cite{sarmento2023gtr}\cite{sarmento2024Guitarphd}, automatic guitar transcription \cite{wiggins2019guitar} \cite{cwitkowitz2023fretnet}, tablature prediction from MIDI \cite{edwards2024midi} and acoustic guitar synthesis from tablatures \cite{kim2024expressive}.

As previously discussed, representing tablatures is difficult as there is no standardised representation like MIDI. Some works use a JSON-like representation to encode information \cite{xi_guitarset_2018} \cite{kehling2014automatic}. DadaGP \cite{sarmento_dadagp_2021} also introduced an easy to parse text-based encoding which uses a simple vocabulary to describe strings, fret, and many guitar-specific expressive techniques.

\subsection{Guitar-Focused Datasets}

\begin{table}[h]
\resizebox{\columnwidth}{!}{%
\begin{tabular}{cccc}
\toprule
\textbf{Dataset} & \textbf{Length (m)} & \textbf{MIDI} & \textbf{Tab} \\ \hline

GAPS \cite{riley2024gaps} & 843 & \cmark & \xmark  \\
\textbf{GOAT [Ours]} & \textbf{354} & \textbf{\cmark} & \textbf{\cmark} \\
IDMT \cite{kehling2014automatic} & 340 & \xmark & \cmark \\
Guitar-TECHS \cite{pedroza2025guitar} & 312 & \cmark & \cmark \\
GuitarSet \cite{xi_guitarset_2018} & 180 & \xmark & \cmark  \\
FrançoisLeduc \cite{riley2024high} & 240 & \cmark & \xmark \\
EGDB \cite{chen2022towards} & 118 & \cmark & \xmark \\
\bottomrule
\end{tabular}%
}
\caption{Guitar-focused datasets in the literature. We report real audio content duration (in minutes), and existence of annotations in MIDI and Tablature format. We consider tablature format to be any kind of annotation containing string and fret information.}
\label{tab:datasets}
\end{table}

To facilitate guitar focused research, a number of datasets has been compiled for various purposes, focusing mainly on the task of AGT, as observable in Table \ref{tab:datasets}. Datasets such as GuitarSet \cite{xi_guitarset_2018}, EGDB \cite{chen2022towards}, GAPS \cite{riley2024gaps} (which focuses exclusively on classical guitar recordings), IDMT-SMT-Guitar \cite{kehling2014automatic} FrançoisLeduc \cite{riley2024high} and Guitar-TECHS \cite{pedroza2025guitar} contain real guitar recordings with annotations of notes, typically in MIDI or tablature-like form.  IDMT-SMT-Audio-Effects \cite{stein2010automatic} focuses on guitar recordings with various audio effects.
SynthTab \cite{zang2024synthtab}  contains an extremely large collection of audio renditions of guitar tracks synthesised using a virtual instrument. While still sounding somewhat realistic, we do not include SynthTab in Table \ref{tab:datasets} due to the synthetic nature of the audio.

\begin{figure*}[h]
    \centering
    \includegraphics[width=\textwidth]{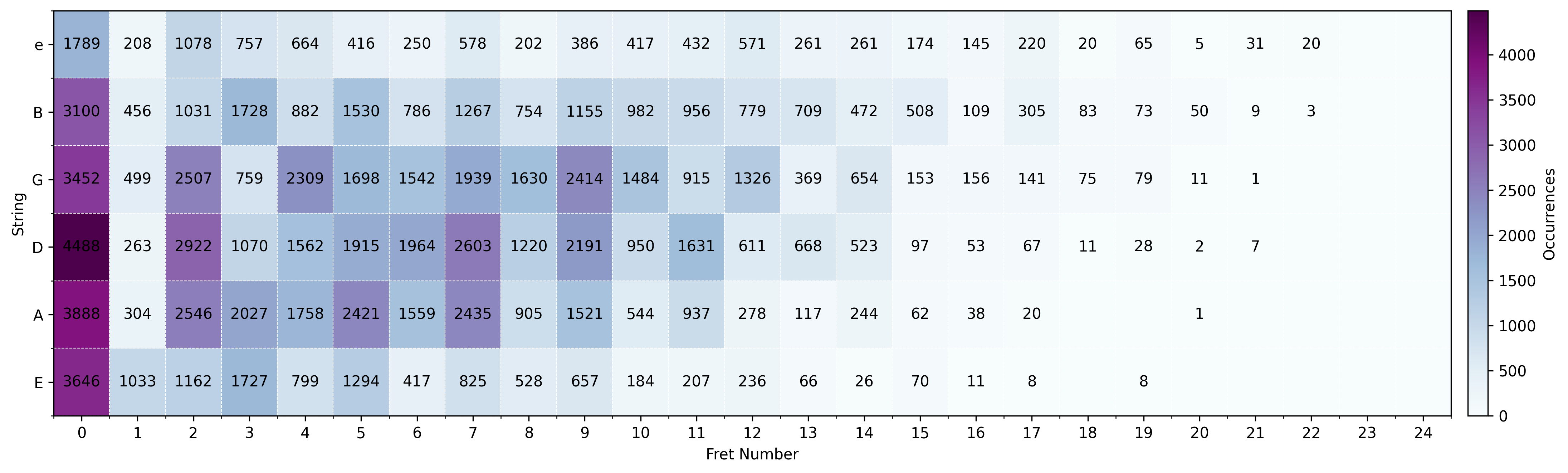}
    \caption{Note distribution over the fretboard (i.e. string and fret positions) of the content in the GOAT dataset.}
    \label{fig:fretboard}
\end{figure*}

\section{Methodology}\label{sec:met}
In order to have as much flexibility as possible with the dataset, it was decided early on that the dataset would have the following qualities: (1) direct input (DI) recordings to allow for post-processing towards tone variety; (2) 44.1 kHz sampling rate to account for high-quality audio; (3) annotations in the form of guitar tablatures to capture as much detail about the playing as possible.

\subsection{Data Collection}
The data was produced by both the main authors and two third party content creators. These creators delivered digital audio workstation (DAW) projects containing a preexisting cover of a popular song and corresponding tablatures, from which the audio and tab pairs were extracted. This allowed us to quickly obtain a large amount of data. Each song was manually checked and aligned against the tablature to ensure that every note was correct between audio and tablature. If multiple guitar parts were present in a tab, they were split off into separate files. In addition to this, we recorded over an hour of additional audio to complement the dataset. This was done by collecting community-created tablatures\footnote{\href{https://www.songsterr.com/}{https://www.songsterr.com/}} and recording the parts in each song exactly as described in the tablature.

\subsection{Data Processing}
In order to protect the work made by the third party creators, all song and artist names were removed from the file names and tablatures. The tablatures were all converted to the DadaGP \cite{sarmento_dadagp_2021} text encoding format due to its popularity in tablature-related tasks. MIDI was also generated from the tablatures using the Guitar Pro MIDI export function. While the MIDI annotations do not encode the expressive techniques included in the tabs, they allow the dataset to be used along with other datasets and models which were built for MIDI annotations, leveraging contributions from the literature. Despite the actual notes played in the audio files having small timing imperfections, the notes in the tablature annotations are all quantized to a grid. The alignment procedure from \cite{riley2024gaps} was used to fine-align the MIDI notes to the performances. Both the fine-aligned and quantized MIDI are provided in the dataset. The tabs were also rendered using the Guitar Pro Realistic Sound Engine (RSE)\footnote{\href{https://www.guitar-pro.com/blog/p/14545-signature-sounds-explained-guitar-pro-7}{https://www.guitar-pro.com/blog/p/14545-signature-sounds-explained-guitar-pro-7}}. virtual instrument to create additional synthetic data. %Finally, to simplify the dataset, we searched through GOAT for any songs that used non-standard tunings and sectioned them off from the main corpus. This data is still annotated in the same way as the rest of the dataset, but filtering them off allows researchers to assume standard tuning for the main part of GOAT.

\subsection{Amplifier Rendering}
\label{sec:amplifier}

The audio in GOAT is recorded from the raw output of an electric guitar pickup, also called direct input or DI. This is not a sound commonly heard in music, as the electric guitar sound most listeners are accustomed to comes from a DI run through a guitar amplifier and guitar speaker cabinet connected in series. We collected DIs in order to maximize the potential of data augmentation via reamping. Reamping is a process in which a DI is run through an amplifier and sometimes a full effects chain to transform it into a desired final sound. This is common in music production when an engineer wants to capture a performance but not commit to a final guitar tone. 

While it would be theoretically possible to reamp GOAT using a variety of real amplifiers and cabinets, this process is time consuming and expensive. Instead, we can use high quality digital amplifier models and impulse responses (IRs) of a guitar cabinet speaker to model the amplifier and speaker cabinet respectively. This approach was used by \cite{chen2022towards} and showed improvement in transcription with the increased tone variety, though the amplifier models used were of low quality and limited to just five total digital amplifiers. We extend this idea by using a higher quality modern amplifier modeling software, as well as significantly increasing the variety in amplifiers and cabinets used. Through this method, we are able to transform the audio in GOAT much closer to the sound of real world data, as well as significantly increase the timbral variety. 

For the amplifier, we used the Neural Amp Modeler\footnote{\href{https://www.neuralampmodeler.com/}{https://www.neuralampmodeler.com/}}(NAM), a high quality digital amplifier plugin. The NAM allows users to capture a `profile' of a guitar amplifier, essentially a snapshot of the amplifier's sound. A large community\footnote{\href{https://tonehunt.org/}{https://tonehunt.org/}} has grown out of users sharing their profiles online. Using publicly available NAM profiles, we curated a dataset of roughly 7,000 profiles from over 1,000 different amplifiers. A profile was chosen at random to render each audio item in the dataset. IRs were used from the NeuralDSP Archetype Nolly\footnote{\href{https://neuraldsp.com/plugins/archetype-nolly}{https://neuraldsp.com/plugins/archetype-nolly}} plugin. This plugin has cabinet IRs from several different microphones, with parameters to control the microphone height and distance from the speaker. The microphone and its parameters were chosen at random for each audio item in the dataset. A send to a room reverb was also enabled at random (with a $25\%$ chance of being enabled) and the send level was randomised between $-10$dB and $0$dB. A python package called Pedalboard\cite{sobot_peter_2023_7817838} was used to render the audio. This process was repeated 5 times over, resulting in 29.5 hours of fully annotated audio. This process can, however, be extended by choosing different post-processing approaches. The reamping code\footnote{\href{https://github.com/JackJamesLoth/GOAT-Dataset}{https://github.com/JackJamesLoth/GOAT-Dataset}} is made publicly available.
\begin{figure*}[t]
    \centering
    \includegraphics[width=\textwidth]{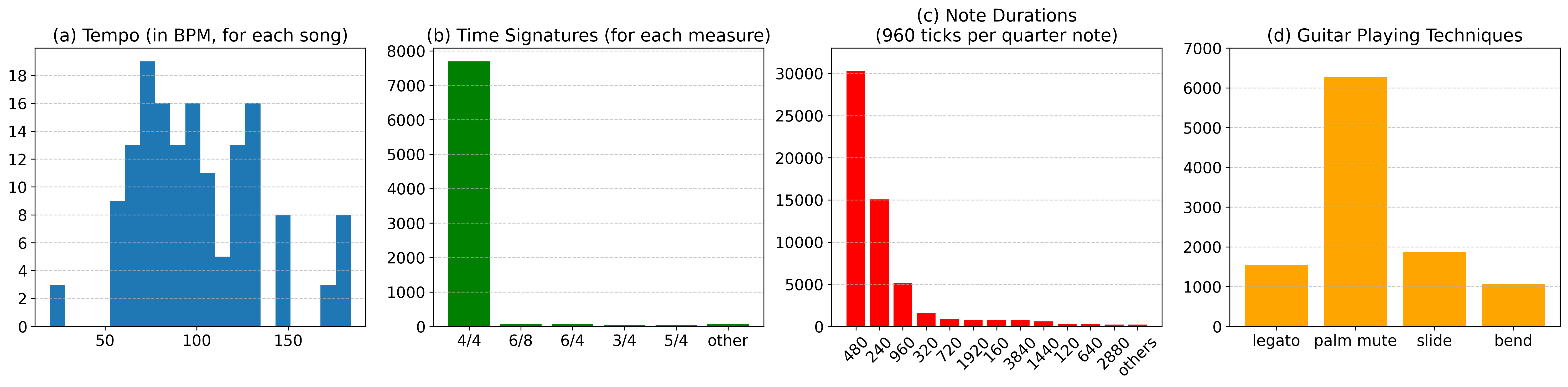}
    \caption{Statistical information about the GOAT dataset. Histograms of tempo (in BPM) per song (a), time signatures per measure (b), most common note durations, in which a value of 960 ticks corresponds to a quarter note (c), and most frequent guitar playing techniques (d).}
    \label{fig:histogram}
\end{figure*}

\vspace{-0.4cm}
\section{Dataset}\label{sec:dataset}

\subsection{Statistical Overview}

The finalised dataset contains 5.17 hours of DI recordings in standard tuning\footnote{Standard tuning uses the open string pitches E, A, D, G, B, e.} and 25.85 hours of amplifier recording, as well as an additional 43.85 minutes in non-standard tuning. This is broken into 153 (standard tuning) and 19 (non-standard tuning) individual dataset items.

For each data point in the GOAT dataset we provide the following: (1)
the DI audio file (in .wav format) and (2) the DI audio rendered through 5 different amplifiers and 5 different cabinets (in .wav format); (3) the corresponding Guitar Pro tablature (in .gp and .gp5 format, the latter to allow for conversion into the DadaGP format), the (4) corresponding DadaGP text encoding of the Guitar Pro annotation (in .txt format), and (5) a rendered version into audio using the RSE virtual instrument in Guitar Pro (in .wav format); finally, (6) the corresponding MIDI and fine-aligned MIDI annotations (in .mid format). A .csv file containing metadata and file paths is also provided.

\begin{table}[h]
    \centering
    \begin{tabular}{ccc}
        \toprule
            \textbf{Guitar} & \textbf{Standard (m)}\ & \textbf{Other (m)} \\
         \hline 
         Stratocaster & $208.82$ & $26.15$ \\
         Les Paul & $6.08$  & $7.36$   \\
         Jazzmaster & $15.62$  & $10.34$ \\
         Strandberg & $79.83$  & - \\
         \bottomrule
    \end{tabular}
    \caption{Breakdown of content of audio (in minutes) by type of guitar, using standard or alternative tunings (e.g. drop-D tunings, half-step or whole-step down-tunings).}
    \label{tab:guitarhours}
\end{table}

Table \ref{tab:guitarhours} reports the content (in terms of minutes) of audio, divided by type of guitar used in the recordings. Overall, it contains 172 distinct audio files, each with all the annotations and metadata described previously.

In Figure \ref{fig:fretboard} we observe a compilation of the note distribution for all the songs in GOAT, from a total of 109,869 total played notes, containing information about the frets and strings used for each. 

\begin{table}[h]
    \centering
    \begin{tabular}{cc}
        \toprule
            \textbf{Chord Type} & \textbf{Number of Instances}\\
         \hline
         w/ 3 notes & $7,174$\\
         w/ 4 notes & $4,260$\\
         w/ 5 notes & $1,065$\\
         w/ 6 notes & $1,039$\\
         \bottomrule
    \end{tabular}
    \caption{Breakdown of number of chords in the GOAT dataset in terms  of number of notes.}
    \label{tab:chords}
\end{table}

Statistics on the chords present in GOAT are shown in Table \ref{tab:chords}. From a total of 13,538 chords in the dataset, there is a prevalence of chords played using only 3 notes. This includes triads (major, minor, diminished, augmented), but mostly ``power chords'', common in the rock/metal sonorities, comprised of the root, and the fifth and octave above.

Figure \ref{fig:histogram} shows histogram breakdowns of the musical content in the dataset. In (a) we observe a distribution of the tempos, in bpm, for each song in the GOAT dataset. In terms of time signatures, in (b) we see a prevalence of 4/4 time signatures in most of the measures of the songs in the dataset. The statistics presented in (c) and (d) leverage the token format of the DadaGP annotations. Inspecting the GOAT dataset via this text-like format, we can produce an histogram of the most common note durations (see (c), in which a quarter note corresponds to 960 ticks, 480 ticks correspond to an eighth note, 240 ticks to a sixteenth note, 320 to an eighth note triplet, 720 ticks to a dotted eighth note, and so on), and the most commonly used guitar playing techniques (d). Regarding the latter, we have highlighted the four most prevalent expressive techniques, but the dataset contains instances of others (e.g. tapping, vibrato), although with a much smaller representation.

\subsection{Distribution}

We distribute the GOAT dataset on the Zenodo\footnote{\href{https://zenodo.org/records/15690894}{https://zenodo.org/records/15690894}} platform. The dataset is made available by request to better control its use for research purposes only.

\subsection{What Is Missing?}

As of this version, there is no information regarding key signature as metadata in GOAT. Finally, akin to what the authors referred to in \cite{sarmento_dadagp_2021}, Guitar Pro (up until its version 5) does not include note velocity information as in MIDI. However, Guitar Pro does represent loudness between notes and musical phrases by using traditional dynamic instructions (e.g. forte, piano). Thus, dynamics are excluded from both the Guitar Pro and MIDI renditions of the GOAT dataset, and they are exclusively contained in the realistic DI recordings. In terms of metadata, it would be interesting to provide a distribution of the type of chords in terms of harmonic quality (e.g. major, minor, seventh) as well as diagrams or fingerings, but that information is currently absent from what GOAT provides.% Finally, the fact that the neck pickup was used in all the recordings hinders the variance of timbral characteristics of the data.

\section{Experiments Using GOAT}
\label{sec:experiments}

In order to showcase the applicability of GOAT, we selected automatic music transcription as a use case. We evaluate GOAT on two tasks, transcription into MIDI format, (here referred to as MIDI transcription), focusing only on capturing note's pitch, duration and velocity information, and a newly proposed task of automatic guitar tablature transcription (AGTT) via DadaGP token prediction.

\subsection{MIDI Transcription}
\label{sec:miditranscription}

While GOAT is not expressly a MIDI-focused dataset, MIDI transcription is still the most common baseline used in automatic guitar transcription (AGT) and is the method researchers will be most familiar with. Despite the previously discussed shortcomings of MIDI, it is still beneficial to evaluate GOAT using the task of MIDI transcription in order to better understand how the dataset compares to other similar datasets.

Following \cite{riley2024gaps}\cite{riley2024high}, we finetune a high-resolution transcription model \cite{kong} pretrained using extensively augmented data \cite{edwards_aug}. Rather than training from scratch, we finetune a pretrained model, as \cite{riley2024gaps, riley2024high} have demonstrated that finetuning yields superior performance. We use the same training setup as \cite{riley2024gaps} where each training data is segmented into 10-second chunks with a hop size of 1 second. During training, we apply data augmentation by randomly shifting the pitch up to $\pm 2$ semitones. 

We train three separate models using GOAT for our experiments. The models were trained on an NVIDIA A5000 GPU until convergence. The first uses only the raw DI recordings (DI). The next uses the dataset reamped once using our methodolgy discussed in Section \ref{sec:amplifier} (AMP). Finally, we train a model using the dataset reamped twice (AMP-XL).  While this does not mean that the model would see more variety in MIDI annotation during training, it does mean that the model will see more timbral variety during training. These models are then evaluated on the test split of GOAT (both DI and AMP) and GuitarSet \cite{xi_guitarset_2018} (which contains only acoustic guitar). We choose to evaluate our model on GuitarSet to assess its zero-shot learning capabilities. This is done in order to test the following hypotheses: \bm{H_1} - data augmentation in the form of guitar amplifier reamping improves the generalizability of a transcription model on real data; \bm{H_2} - reamping the dataset multiple times to give additional timbral variety further improves the generalizability of a transcription model on real data. The MIDI transcription results are found in Table \ref{tab:midiresults}.

\begin{table}
    \centering
    \begin{adjustbox}
    {width=\columnwidth} 
    \begin{tabular}{ccS[table-format=1.3]S[table-format=1.3]S[table-format=1.3]}
        \toprule
        \textbf{Data} & \textbf{Test Split} & $\bm{P}$ & $\bm{R}$ & $\bm{F}$\\
        \midrule
         & DI & 0.8470090815161626& 0.7978721496665799&0.8204153813633766\\
        DI & AMP &0.7705574552258332&0.6773443856679099&0.7141396209991608\\
        & GuitarSet &0.8794249922315108&0.8490797236872035&0.8600160178935202\\
        \midrule
        & DI &0.8528920968600798&0.8359556403112557&0.8425736719240288\\
        AMP & AMP &0.8236783419392569 &0.781367260354084&0.8003546550481061\\
        & GuitarSet &0.881786612045042&0.814721879036653&0.8426039800630928\\
        \midrule
        & DI &0.831&0.809&0.818\\
        AMP-XL & AMP &0.8212&0.7964&0.8064\\
        & GuitarSet &0.8723&0.8083&0.8346\\
        \bottomrule
    \end{tabular}
    \end{adjustbox}
    \caption{MIDI transcription results. Metrics used are note-level precision ($P$), recall ($R$) and $F_1$ score ($F$).}
    \label{tab:midiresults}
\end{table}

\subsubsection{Discussion}

All three models in general achieve fairly competitive results on the test datasets \cite{riley2024high}. The AMP model is able to obtain comparable results to the DI model on the DI and GuitarSet test split, while significantly outperforming the DI model on the AMP test split, the latter being the closest to a real world transcription scenario concerning electric guitar. This is strong evidence for $\bm{H}_1$, suggesting that the amplifier data augmentation allows the model to generalize to different amplifier timbres without sacrificing quality in the case of DI or acoustic guitar. This improvement also surpasses the gains observed between models trained with DI and amplifier data in \cite{chen2022towards}, further validating the effectiveness of our data augmentation strategy. Notably, our model trained solely on DI data achieves comparable transcription performance on GuitarSet's test set \cite{riley2024high}, further validating GOAT’s effectiveness in training AGT systems.

Interestingly, the AMP-XL model performs worse than both the DI and AMP models on the DI and GuitarSet test splits, while performing slightly better than AMP on the AMP test split. This seems to suggest that the additional timbre variety is only helping generalisation among amplifier timbres, while hurting generalisation among the unseen DI and acoustic guitar timbres. However, we posit that this still shows some evidence for $\bm{H}_2$ given that the AMP test split is the closest to a real world electric guitar transcription use case.

\subsection{Tablature Transcription with Whisper}

Given our choice of Guitar Pro as the annotation format in GOAT, an obvious choice of task to evaluate the dataset is AGTT. While this task has seen some previous work \cite{wiggins2019guitar} \cite{cwitkowitz2023fretnet}, it is still conducted in a manner similar to MIDI transcription by predicting string and fret activations at small time scales. We formulate the problem differently and propose a new task by instead using the DadaGP annotations as plain text and treating the problem as an audio-to-text problem. This approach means that we are not predicting exact note timings, but rather the more general sheet music-like interpretation of the input audio.

For this task, we fine-tune Whisper \cite{radford2023robust}, a large foundation model trained for automatic speech recognition. The audio in GOAT was split into 15 second chunks, as Whisper has a maximum audio input length of 30 seconds. Whisper also has a maximum sequence length of 448, so any chunks whose annotations exceeded 448 tokens were excluded. This meant we were only able to use roughly 4.65 hours of GOAT. A custom tokenizer was also created so that the model could correctly interpret and predict DadaGP tokens. The vocabulary used in DadaGP was simplified, removing tokens relating to artist and genre and simplifying note tokens by removing timbre information.

In our experiments, we again train three different models using the same data setup as in Section \ref{sec:miditranscription}. The models were trained on an NVIDIA A100 GPU until convergence. We use token accuracy and word error rate (WER) to evaluate the performance. Preliminary results from the best performing epochs are presented in Table \ref{tab:whisperresults}.

\begin{table}[]
    \centering
    \begin{tabular}{cS[table-format=1.3]S[table-format=1.3]S[table-format=1.3]}
    \toprule
        \textbf{Data} & {\textbf{Loss}} & {\textbf{Accuracy}} & {\textbf{WER}}  \\
        \midrule
        DI & 1.787020 & 0.682151 & 0.514608 \\
        AMP & 2.065203 & 0.681347 & 0.547430\\
        AMP-XL & 2.285448 & 0.692201 & 0.518440\\
        \bottomrule
    \end{tabular}
    \caption{Evaluation results on the test split of GOAT from the best runs of the fine-tuned Whipser model.}
    \label{tab:whisperresults}
\end{table}

\subsubsection{Discussion}

The transcription results ultimately did not work well, though they do show some promise. DadaGP is a highly structured text encoding, and the biggest initial hurdle is learning this structure. For example, in a DadaGP encoding, notes are specified through one or more \texttt{note} tokens, followed by a \texttt{wait} token. Given the abundance of \texttt{wait} tokens in DadaGP encodings, it is possible for the model to simply learn to output only these tokens to optimise the loss function. While example outputs of the fine-tuned models to show this behavior in parts of the prediction, the first half of the token predictions still tend to be semi-structured DadaGP encodings. Common note groups (e.g simple chords) are predicted. However, these ultimately did not correlate with the notes in the ground-truth encodings. While the accuracy appears to be somwhat high, this is likely misleading due to the prelevance of \texttt{wait} tokens in both the dataset and model outputs.

It seems likely that more fine-tuning and data is needed to get a workable transcription model. Training using the AMP-XL data resulted in a small improvement to the accuracy and WER, suggesting that additional data could be beneficial. A pretraining step using synthetic data, such as SynthTab \cite{zang2024synthtab} for example, could be a feasible way to collect enough data to get the model to properly learn the DadaGP structure. This could then be followed by a further fine-tuning step on GOAT to expose the model to a wide range of timbres and real guitar playing. However, we leave these improvements for future work. We include these limited results in order to show the potential of this approach and showcase the benefit of a dataset like GOAT, which is unique in being the only non-synthetic dataset of paired guitar audio and tablatures. 

\section{Prospective Use Cases}\label{sec:app}

We envision GOAT as a general purpose dataset that can be useful for a number of different tasks. While we propse these tasks as the starting point of work with this dataset, we hope that the dataset finds uses in other unique and interesting tasks as well. 

\subsection{Automatic Guitar Transcription}
Containing both MIDI and tablature annotations of the source audio, GOAT lends itself to AGT very well. Section \ref{sec:experiments} details preliminary results for MIDI-based transcription and a newly proposed task of DadaGP token prediction for tablature transcription. We leave improvement on these baselines as future work.

\subsection{Realistic Guitar Synthesis}

GOAT provides numerous examples of different notes and playing techniques, with annotations for all of them. The data can be used to help train or fine-tune popular models such as RAVE \cite{caillon_rave_2021} and DDSP \cite{engel_ddsp_2020}. It is also suitable for MIDI synthesis models such as \cite{maman2024performance} or even novel tasks such as synthesis from guitar tablatures.

%GOAT provides numerous examples of different notes and playing techniques, with annotations for all of them. The amount of data is suitable for popular models such as RAVE \cite{caillon_rave_2021} and DDSP \cite{engel_ddsp_2020}. Furthermore, a companion paper\footnote{Paper under review - citation will be added pending acceptance.} to this one demonstrates the applicability of GOAT on realistic guitar synthesis from tablatures using Flow Matching \cite{lipman2022flow} approaches.

\subsection{Automatic Guitar Playing Technique Identification}

Automatic guitar playing technique identification is another task that we believe can benefit from the availability of GOAT. For example, the authors in \cite{DHooge2023ModelingBends}, propose ways of modelling bends in guitar tablatures. Moreover, while datasets exist for this type of work \cite{chen2015electric}, GOAT has significantly more audio and contains both playing technique and note annotations. We hypothesise that this could potentially be combined with AGTT, further improving the capabilities of guitar transcription models.

\subsection{Effect/Distortion Removal}

The clean DI audio and amplifier pairs provide numerous examples for effect/distortion removal techniques \cite{rice2023general} \cite{imort2022distortion}. The data augmentation techniques can also be extended to other common effects (e.g phaser, chorus).

\section{Conclusion}

We present GOAT, a vast dataset containing pairs of real guitar audio and guitar tablature annotations. This dataset is unique among guitar-focused datasets in that it uses digital representations of tablatures as the annotation for real audio, an annotation format which is well suited to capture guitar-specific metadata. A data augmentation strategy which offers a large amount of tonal variety is presented. We validate the use of the dataset and the augmentation strategy through a standard literature-based MIDI transcription experiment. The results show competitive performance as well as considerable improvement to generalisation when using data augmentation. We also show preliminary steps towards AGTT via DadaGP tokens, a task uniquely suited to GOAT. Finally, we propose prospective interesting applications and paths of research utilising GOAT including transcription, synthesis, playing technique identification and effect removal tasks. We hope that this dataset helps researchers further improve guitar-focused MIR research.

\section{Ethical Statement}

We declare that, in the creation of the GOAT dataset, careful consideration was undertaken by the authors to ensure that the process follows the ethical guidelines supported by the ISMIR community. In this regard, we inform that (1) the content creators were made aware of the prospective applications and use cases for the data they provided, that (2) they agreed to such applications, and that (3) they were duly compensated for the data. However, due to the fact that the GOAT dataset contains covers (i.e. subjective renditions, performances and annotations/transcriptions) of popular songs, both in recorded audio and tablature formats, and because we do not exclusively own the copyrights for some of its content, we intend to make the dataset available for research purposes only, upon request. 

%Finally, in an attempt to reduce carbon footprint from training procedures of AI-based models, we intent to further release the code and checkpoints for the AGT approaches described in the paper.
 \section{Acknowledgements}
This work is supported by the EPSRC UKRI Centre for Doctoral Training in Artificial Intelligence and Music (Grant no. EP/S022694/1) as well as UKRI Innovate UK (Project no. 10102804).

\bibliography{main}

\end{document}